Structural properties and superconductivity of $SrFe_2As_{2-x}P_x$ (0.0≤x≤1.0) and $CaFe_2As_{2-y}P_y$ (0.0≤y≤0.3)

H. L. Shi, H. X. Yang, H. F. Tian, J. B. Lu, Z. W. Wang, Y. B. Qin, Y. J. Song, and J. Q. Li*

Beijing National Laboratory for Condensed Matter Physics, Institute of Physics,

Chinese Academy of Sciences, Beijing 100190, People's Republic of China.

**Abstract**

The $SrFe_2As_{2-x}P_x$ (0.0≤x≤1.0) and $CaFe_2As_{2-y}P_y$ (0.0≤y≤0.3) materials were prepared by a solid state reaction method. X-ray diffraction measurements indicate the single-phase samples can be successfully obtained for $SrFe_2As_{2-x}P_x$ (0.0≤x≤0.8) and $CaFe_2As_{2-y}P_y$ (0.0≤y≤0.3) samples. Clear contraction of the lattice parameters are clearly determined due to the relatively smaller P ions substation for As. The SDW instability associated with tetragonal to orthorhombic phase transition is suppressed visibly in both systems following with the increase of P contents. The highest superconducting transitions are respectively observed at about 27 K in $SrFe_2As_{1.3}P_{0.7}$ and at about 13 K in $CaFe_2As_{1.7}P_{0.3}$.

**Introduction**

Following the discovery of superconductivity at $T_c \approx 26$ K in LaFeAsO(F) [1], a variety of isostructural compounds were synthesized under different conditions, such as RTMPO (R=rare earth metal, TM = transition metals) [3-7], RTMAsO [8-17], and SrFeAsF [20] (so called the 1111 type). These materials could exhibit numerous interesting physical properties, e.g. ferromagnetism (FM), antiferromagnetism (AFM), spin density waves (SDW), and superconductivity. Similar structural and physical phenomena have been also observed in the $AFe_2As_2$ materials (A=Ca, Sr, Ba, Eu, often called 122 phase) [14-19]. Electron or hole doping [20-22] in a umber of systems can suppress the SDW and result in superconductivity in both the 1111 phase and 122 phase. Recent experimental investigations revealed that the P substation of As in $BaFe_2As_2$ could also induce superconductivity at about 31 K [24]. Theoretic calculation [25] suggests that Phosphorous doping does no significantly change the electron density, rather than evenly influence on the location of hybridized states, band width and the topography of Fermi surface. Therefore, the influence of P substitution for As on the physical properties, especially the superconductivity in related system, is still a open issue. In this paper, a series of single phase polycrystals of $SrFe_2As_{2-x}P_x$ and $CaFe_2As_{2-y}P_y$ samples have been prepared, the highest superconducting critical temperature $T_c$ is obtained at about 27 K in $SrFe_2As_{1.3}P_{0.7}$ and 13 K in $CaFe_2As_{1.7}P_{0.3}$ materials, respectively. We also noted that local structural properties of FeAs layer is a critical factor which significantly influences crystal structure and transportation properties, as reported in previous literature [25].

**Experimental methods**

Polycrystalline samples of SrFe$_2$As$_{2-x}$P$_x$ (0.0≤x≤1.0) and CaFe$_2$As$_{2-y}$P$_y$ (0.0≤y≤0.3) were synthesized by solid-state reactions as reported in the previous literatures [16, 23]. The starting materials of Ca(Sr)Fe$_2$As$_2$ and Ca(Sr)Fe$_2$P$_2$ were prepared by using small Ca or Sr chunks and the high quality of Fe, As and P powders. The mixtures with desired compositions are pressed into pellets, placed in a small alumina crucible, and then sealed in a silica tube, each tube was filled with 1/3 Argon gas. The tube was heated at 800℃ in 24 h, then cooled down to the room temperature. The obtained resultants were grinded as starting materials. SrFe$_2$As$_{2-x}$P$_x$ and CaFe$_2$As$_{2-y}$P$_y$ materials were prepared using these starting materials, grinded and pelleted, and sealed in vacuum tubes which were heated at 850 ℃ for 48 h and cooled down to room temperature for 12 h. X-ray diffraction (XRD) measurements on all samples were carried out on a diffractometer in the Bragg–Brentano geometry using Cu $K_a$ radiation. Magnetization measurements as a function of temperature were performed using a commercial Quantum Design SQUID. The resistivity (R) as a function of temperature was measured by a standard four-point probe technique.

**Results and discussion**

1.1 Structural and physical properties of SrFe$_2$As$_{2-x}$P$_x$

Figure 1 (a) shows the powder x-ray diffraction (XRD) patterns of the SrFe$_2$As$_{2-x}$P$_x$ (0.0≤x≤1.0) materials, the experimental results for the samples with x<0.3 could be well indexed on the basis of tetragonal ThCr$_2$Si$_2$-type structure with the space group I4/mmm (No. 129, Z=2). An impurity phase as marked as by an arrow in the diffraction pattern appear progressively following with the increase of P content in addition to the main 122 phase; samples with x≥0.8 are found not to be stable and often decompose in the moist air in several days. Figure 1 (b) and (c) show the lattice parameters as the function of the P content, illustrating the c-axis parameter and volume of unit cell decrease gradually with the increase of P content due to the relatively smaller atom radium of Phosphorous (98 pm) in comparison with that of Arsenic (114 pm). This fact suggests an effective chemical substitution of P for As in our experiments. The experimental data for lattice parameters are list in the table 1.

Figure 2 (a) presents the temperature-dependent resistivity for the $SrFe_2As_{2-x}P_x$ samples, each R-T curve is normalized by its maximum value for facilitating the illustration of resistivity changes accompanying with the SDW stability and superconducting transition. For the parent phase the pronounced resistivity anomaly at 220 K is considered arising from the structural phase transition from tetragonal (I4/mmm) to orthorhombic phase (F mmm) in association with SDW instability. The SDW anomaly is significantly suppressed with P substitution for As and moves to the lower temperature. For instance, the SDW in the x=0.5 sample becomes almost invisible and shows a clear down-turn in resistivity at about 20 K, the x=0.6 sample has a linear resistivity in the temperature range from 300K down to 30K and drastically drop of resistivity undergoes at about 27 K corresponding a superconducting transition. Further increase of the P content leads to the appearance of impurity phase in the samples and no obvious improvement of $T_c$ is observed. Our careful analysis of $SrFe_2As_{2-x}P_x$ suggests that the optical $T_{c\text{-onset}}$ is about 27 K, $T_{c\text{-zero}}$ is about 23 K for the P content at around 0.8. Fig. 2b shows the experimental results of magnetization measurement demonstrates a clear superconducting transition just below 30 K for the x=0.7 sample. It should be noted that for the samples with x<0.6 there is an obvious resistivity down-turn at about 18 K, similar behavior was reported in the $EuFe_2As_2$ sample [17]. The origin of this anomaly is still a problem in discussion.

Electron diagram T-x for the $SrFe_2As_{2-x}P_x$ samples is plotted in the figure 3, where the onset $T_{c\text{-onset}}$ and $T_{SDW}$ were estimated from resistivity measurements using the intersection point of two lines method, and Phosphorous content is adopted the nominal data although a slightly deviation presents in the samples with x>0.7. Comparison with the Co substitution of $SrFe_{2-x}Co_xAs_2$ materials, the suppression of SDW in $SrFe_2As_{2-x}P_x$ is rather moderate, since P substitution for As does not directly on the Fe site which plays a critical role for the superconductivity in present system.

1.2 Structural and physical properties of $CaFe_2As_{2-y}P_y$

Using the similar preparation route we have successfully synthesized $CaFe_2As_{2-y}P_y$ (y=0.0～0.3) polycrystals. The higher Phosphorous content samples (y=0.3～1.0) are also prepared, but they are rather unstable and prone to decomposition in the moist air just like the case of $SrFe_2As_{2-x}P_x$ with x>0.8. Figure 4 (a) presents the powder X-ray diffraction patterns, in which the main reflection peaks can be well indexed with the tetragonal phase I4/mmm except that the additional peaks of impurity phase arise between 112 and 105 peaks for the samples of y⩾0.10. Figure 4 (b) and (c) shows the lattice parameters as the function of Phosphorous content, indicating the visible contraction of lattice parameters due to the relatively small size of P ion in comparison with As. Details of lattice parameters are listed in the table 1.

Figure 5 (a) shows the temperature dependent of the normalized resistivity curve, a small amount of P (y=0.025) could drastically suppress the SDW from 220K down to $T_{SDW}$ = 120 K, while a visible down-turn also appears at about 13 K. When Phosphorous content is up to y=0.075 this SDW phase transition almost disappear and a superconducting transition occurs at about 13 K, but zero resistivity is not obtained. Further increasing Phosphorous content there is little space to improve $T_c$. A magnetization measurement indicates that a significantly diamagnetic signal is recorded at about 11 K, suggesting this resistivity drop corresponds to a superconducting transition. However, notably ferromagnetic background signal is often observed, which possibly results from the impurity phase and leads to non-zero residual resistivity below the superconducting transition.

Previous theoretical investigations on the $BaFe_2As_2$ [25] suggest that the Fermi surface is evidently influenced by P substitution on the As site. Substitution of P for As into FeAs layer could not yield additional electrons or holes into the system, but the visible changes of local structures can be observed, such as the As/P ionic positions in the unit cell, ($Z_{As}$ and $Z_P$) [25], these alternations can evidently influence the bandwidth, location of hybridized states and topography of Fermi surface. Structural investigations of the $SrFe_2As(P)_2$ and $CaFe_2As(P)_2$ materials show that $Z_{As}$ and $Z_P$ are 0.3612 [26] and 0.35218 [27], 0.36642 [26] and 0.36433 [27], respectively. The changes of $Z_{As/P}$ of $SrFe_2As(P)_2$ (0.00902) is almost five times larger than that of $CaFe_2As(P)_2$ (0.00209), thus substitution of P for As in the $SrFe_2As(P)_2$ system could induce notable changes of structural and physical properties.

In conclusion, we have successfully prepared a series of $SrFe_2As_{2-x}P_x$ (0.0≤x≤1.0) and $CaFe_2As_{2-y}P_y$ (0.0≤y≤0.3) samples using the conventional solid-state reaction method. Continuous reduction of lattice parameters with increasing P content show that P atoms are effectively substituted for As in our samples. Certain samples with relative high P content are rather unstable and often decompose into powders in the moist air in a few days. The resistivity anomaly associated with the SDW instability is suppressed following with the increase of P content, clear superconducting transitions were observed in $SrFe_2As_{1.3}P_{0.7}$ at 27K and $CaFe_2As_{1.7}P_{0.3}$ at 13K.

**Acknowledgments**

We would thank W. W Huang for dc-magnetization measurements. This work is supported by the National Science Foundation of China, the Knowledge Innovation Project of the Chinese Academy of Sciences, and the 973 projects of the Ministry of Science and Technology of China.

Figure captions

Figure 1: (a) Powder X-ray diffraction patterns of SrFe$_2$As$_{2-x}$P$_x$ (x=0～1.0), where peaks of impurity phase are strengthened as increase Phosphorous contents, which is marked by an arrow; (b) and (c) Lattice parameters as a function of Phosphorous content.

Figure 2: (a) Resistivity as a function of temperature for SrFe$_2$As$_{2-x}$P$_x$ samples, demonstrating that SDW is gradually suppressed as increase Phosphorous content, where the critical temperature is marked as arrow. The inset shows the exaggerated curve of low temperature part. (b) Molar magnetic susceptibility for x=0.7 samples, where Tc-onset is about 30 K.

Figure 3: The electronic phase diagram SrFe$_2$As$_{2-x}$P$_x$ samples, showing the structural and superconducting transitions as the function of Phosphorous content, all the transition temperatures are determined from resistivity measurement using the intersection point of two lines method.

Figure 4: (a) Powder X-ray diffraction patterns of CaFe$_2$As$_{2-y}$P$_y$ samples (y=0～0.3), where impurity phase is marked as by an arrow; (b) and (c) Lattice parameters as a function of Phosphorous content.

Figure 5: (a) Resistivity as a function of temperature for CaFe$_2$As$_{2-y}$P$_y$ samples, where the critical temperature is marked by arrows; (b) Molar magnetic susceptibility for y=0.075 samples.

Table 1: Lattice parameters a, c, c/a ratio and unit cell volume of SrFe$_2$As$_{2-x}$P$_x$ and CaFe$_2$As$_{2-y}$P$_y$ samples.

Table 2: Lattice parameters and Z$_{As/P}$ for typical 122 materials.

Fig. 1

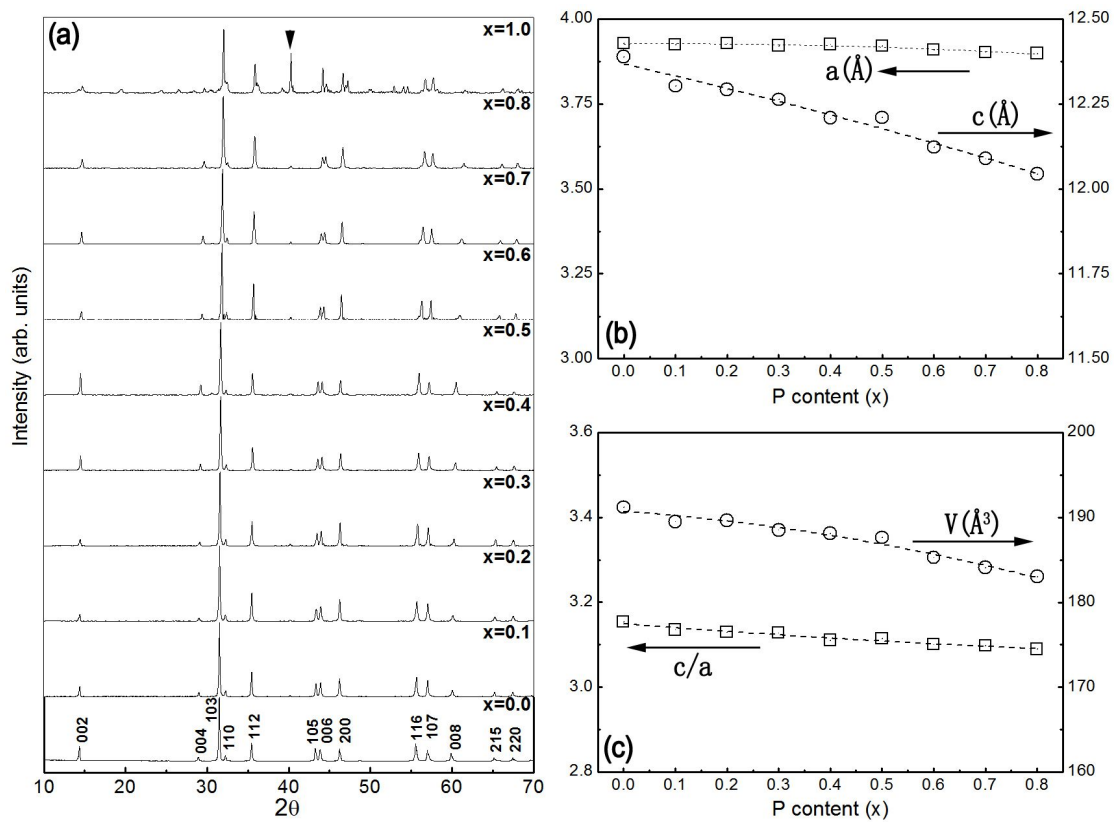

Fig. 2

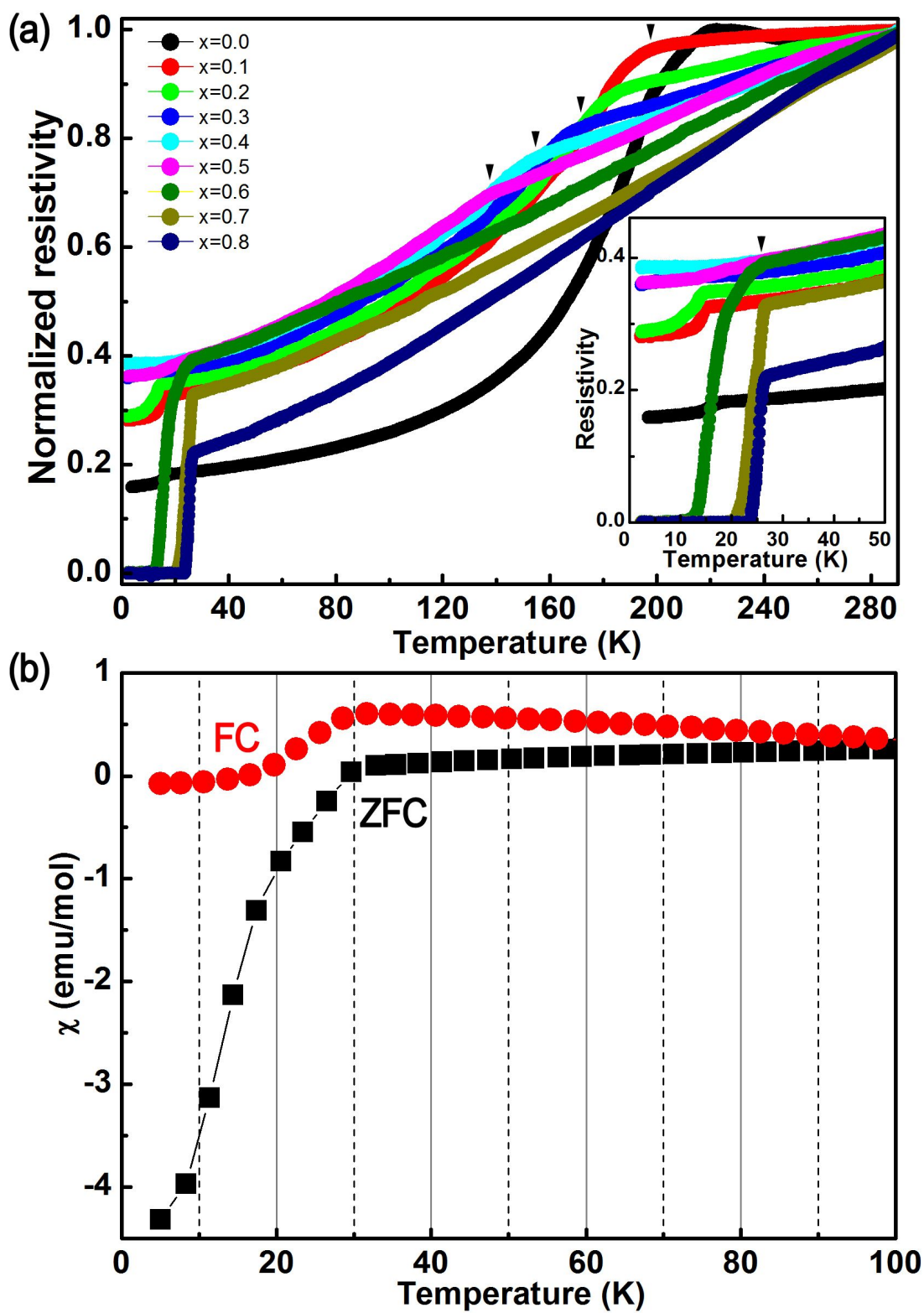

Fig. 3

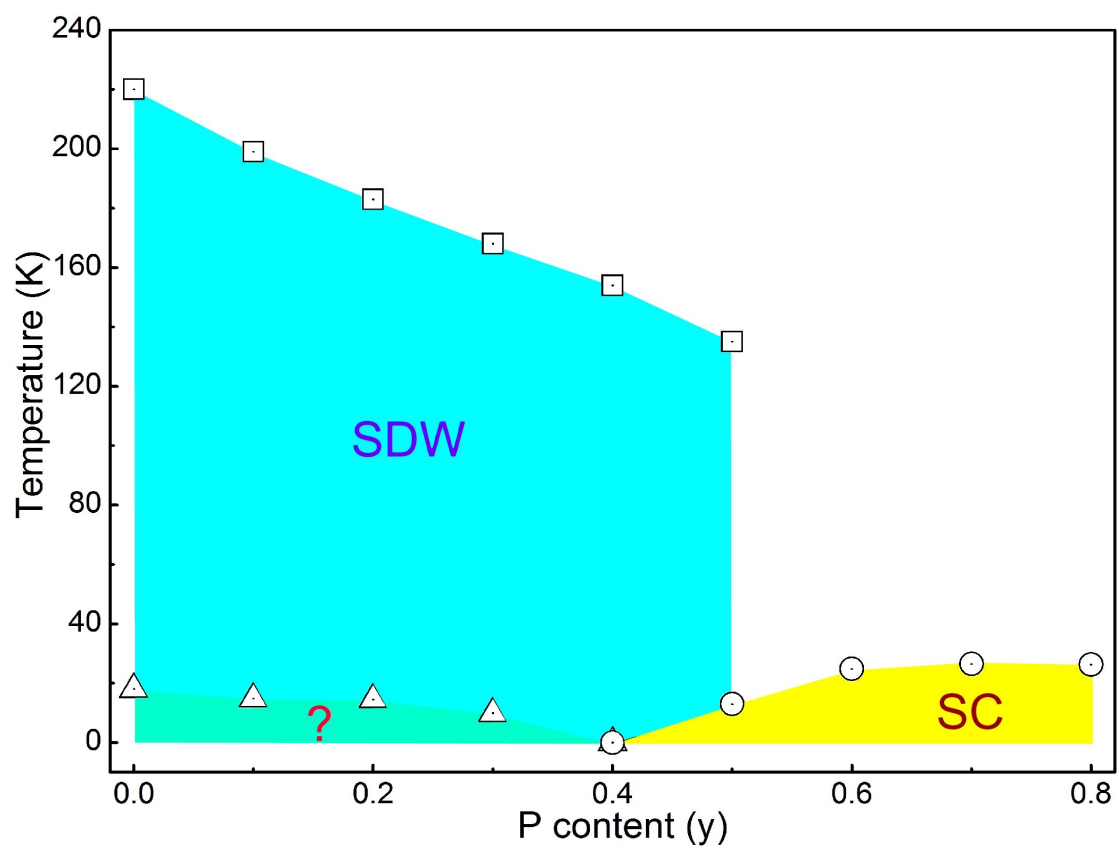

Fig. 4

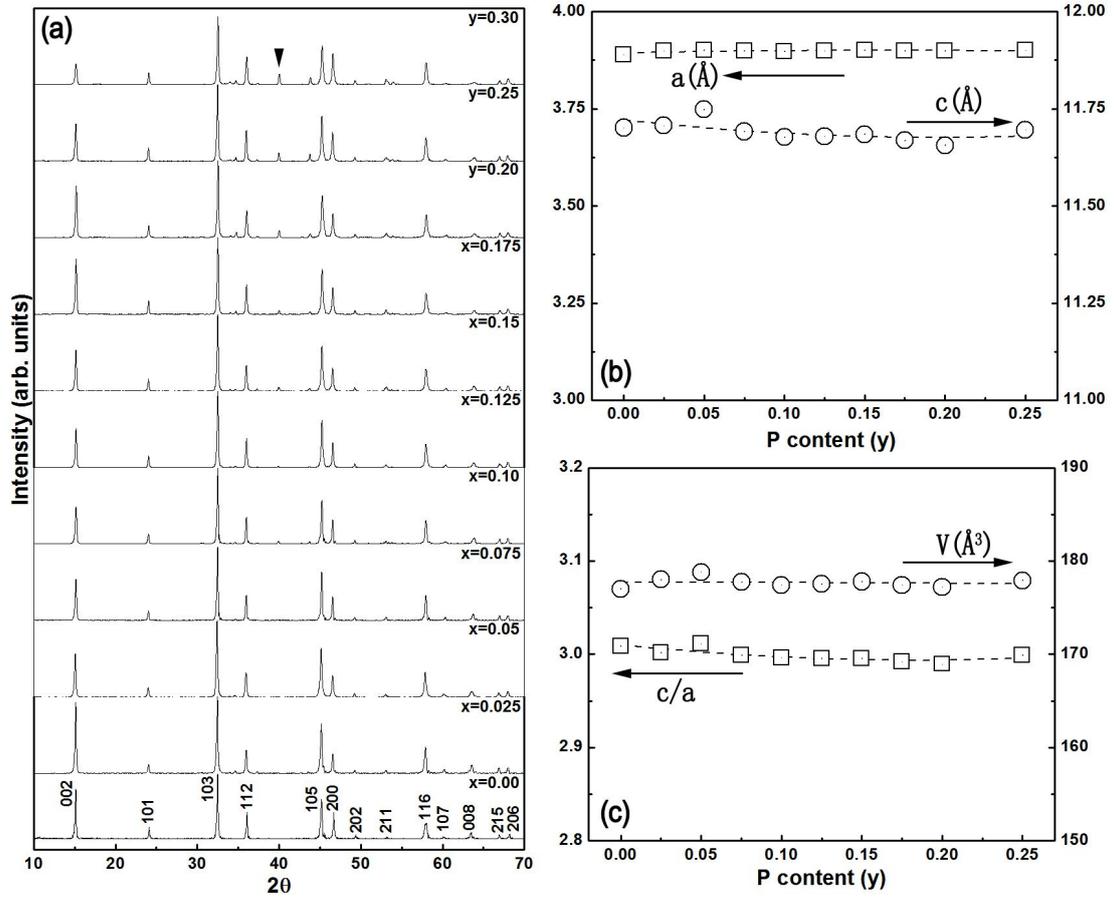

Fig. 5

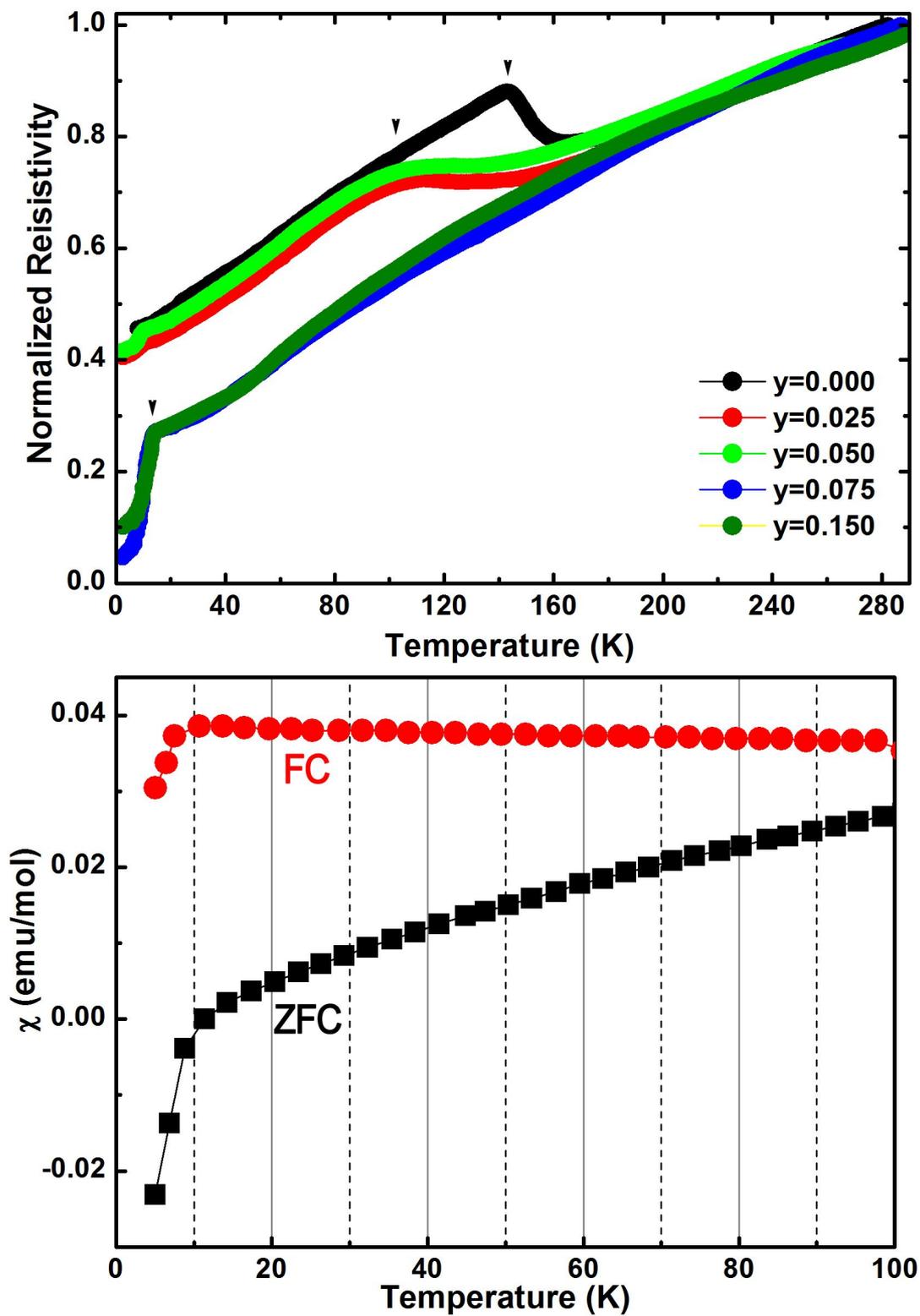

Table 1

| sample | content(x/y) | a (Å) | c (Å) | c/a | V (Å$^3$) |
|---|---|---|---|---|---|
| | 0.0 | 3.928(4) | 12.388(8) | 3.1536(5) | 191.188(0) |
| | 0.1 | 3.924(6) | 12.303(2) | 3.1348(9) | 189.499(9) |
| | 0.2 | 3.927 (6) | 12.292(3) | 3.1297(2) | 189.621(5) |
| | 0.3 | 3.920 (9) | 12.262(7) | 3.1275(2) | 188.520(1) |
| SrFe$_2$As$_{2-x}$P$_x$ | 0.4 | 3.925 (2) | 12.208(8) | 3.1103(6) | 188.103(4) |
| | 0.5 | 3.920(0) | 12.210(0) | 3.1148(0) | 187.623(7) |
| | 0.6 | 3.909 (3) | 12.122(7) | 3.1009(9) | 185.266(7) |
| | 0.7 | 3.902 (2) | 12.089(1) | 3.0980(2) | 184.082(7) |
| | 0.8 | 3.898 (3) | 12.043(3) | 3.0893(7) | 183.018(9) |
| | 0.000 | 3.889(3) | 11.701(2) | 3.0085(6) | 177.001(0) |
| | 0.025 | 3.899(6) | 11.706(3) | 3.0019(2) | 178.016(3) |
| | 0.050 | 3.901(3) | 11.748(1) | 3.0113(3) | 178.807(8) |
| | 0.075 | 3.898(8) | 11.691(5) | 2.9987(4) | 177.718(3) |
| | 0.100 | 3.897(9) | 11.677(3) | 2.9957(9) | 177.420(5) |
| CaFe$_2$As$_{2-y}$P$_y$ | 0.125 | 3.898(8) | 11.679(2) | 2.9955(9) | 177.531(3) |
| | 0.150 | 3.900(7) | 11.684(7) | 2.9955(4) | 177.788(1) |
| | 0.175 | 3.899(4) | 11.66(8) | 2.9922(6) | 177.415(7) |
| | 0.200 | 3.898(9) | 11.656(3) | 2.9896(4) | 177.192(3) |
| | 0.250 | 3.900(1) | 11.695(7) | 2.9988(2) | 177.900(7) |

Table 2

| sample | a(Å) | c (Å) | $z_{As}$ | sample | a(Å) | c (Å) | $z_P$ |
| --- | --- | --- | --- | --- | --- | --- | --- |
| [26]CaFe$_2$As$_2$ | 3.890(8) | 11.727(8) | 0.36642 | [27]CaFe$_2$P$_2$ | 3.642(4) | 9.485(1) | 0.36433 |
| [26]SrFe$_2$As$_2$ | 3.921(8) | 12.337(3) | 0.3612 | [27]SrFe$_2$P$_2$ | 3.825(1) | 11.612(1) | 0.35218 |
| [16]BaFe$_2$As$_2$ | 3.957(0) | 12.968(5) | 0.3545 | [27]BaFe$_2$P$_2$ | 3.840(1) | 12.442(1) | 0.34564 |